\documentclass{PoS}

\title{Structure of the fermion propagator in QED$_{3}$}

\ShortTitle{Structure of the fermion propagator in QED$_{3}$}

\author{\speaker{Yuichi Hoshino}\\
	Kushiro National College of
        Technology,\\
        Otanoshike Nishi 2-32-1,Kushiro City,Hokkaido 084,Japan\\
        E-mail: \email{hoshinor@ippan.kushiro-ct.ac.jp}}

\abstract{The Minkowski structure of the Fermion propagator in $QED_{2+1}$ is evaluated using
a dispersion like method. Including a massive fermion loop to the photon spectral
function, there is no infrared divergences in the $d=-1 $ gauge.  Screening
effects suppress the chiral order parmeter for small numbers of fermion
flavour $N$. We evaluate $\langle \overline{\psi} \psi \rangle $ as a
function of the coupling constant $c$ and $N$. There exists a region 
where chiral symmetry is dynamically broken.}

\FullConference{LIGHT CONE 2008 Relativistic Nuclear and Particle Physics\\
		 July 7-11  2008\\
		 Mulhouse, France}
		 
\begin{document}

\message{ !name(skeleton.tex) !offset(-3) }

\section{Introduction}

It has been known that dynamical mass generation and chiral symmetry
breaking occur in QED$_{3}$ by solving Dyson-Schwinger (D-S) equation or
by computer simulation on a lattice \cite{Kogut,Appelquist,Maris,HM}. Including massless fermion loop to the photon
vacuum polarization, critical behaviour has been studied which was first
pointed out by Appelquist et.al \cite{Appelquist}. The results have not been conclusive for
the existence of critical numbers of flavours. In this work we adopt the
alternative approach to the phase structure of QED$_{3}$ by a dispersion like
method. We evaluate the spectral function for fermion. Assuming physical mass,
we examine the infrared behaviour with a soft photon cloud. As a result, we
obtain a gauge invariant anomalous dimension and a mass shift in the lowest
spectral function in the quenched approximation. Full spectral function is given by
exponentiation of the lowest one. If we include the massive fermion loop to the
spectral function of photon, there is no infrared divergences in the $d=-1$
gauge. When the anomalous dimension equals to one, the propagator behaves as
$1/p^{4}$ at high energy and dynamical chiral symmetry breaking takes place. The
screening effects are large to reduce renormalization constant $Z_{3}^{-1}$
for smal numbers of flavour $N$. We study chiral order parameter $\left\langle
\overline{\psi}\psi\right\rangle $ as a function of $N$ and the coupling
constant $c.$ We show that there is a region in which chiral symmetry is
dynamically broken.

\section{Dispersion approach}
The spectral function of a fermion \cite{Jackiw,Bob} is defined by
\begin{eqnarray}
S_{F}(x)  &  =&S_{F}^{0}(x)\exp(F(x))\\
S_{F}^{0}(x)  &  =& -(i\gamma\cdot\partial+m)\frac{\exp(-m\sqrt{x^{2}})}%
{4\pi\sqrt{x^{2}}},
\end{eqnarray}
where $F$ is an $O(e^{2})$ matrix element $|T_{1}|^{2}$ for the process
electron($p)\rightarrow$electron($r)+$photon($k)$ given by
\begin{equation}
F=\frac{-e^{2}}{i}\int\frac{d^{3}k}{(2\pi)^{2}}\exp(ikx)\theta(k_{0}%
)\delta(k^{2}-\mu^{2})[\frac{m^{2}}{(r\cdot k)^{2}}+\frac{1}{r\cdot k}%
+\frac{(d-1)}{k^{2}}].
\end{equation}
It is well known that the function $F$ has linear and logarithmic infrared
divergences with respect to $\mu$ where $\mu$ is a bare photon mass$.$ Here we
notice the following
\begin{enumerate}
\item $\exp(F)$ includes all infrared divergences$.$
\item The quenched propagator has linear and logarithmic infrared divergences.
\end{enumerate}
Linear divergence is absent in a special gauge. To remove
logarithmic divergences we include a massive fermion loop. The photon
propagator is then written in spectral form
\begin{equation}
D_{F}(p)=\int_{0}^{\infty}\frac{\rho_{\gamma}(\mu^{2})d\mu^{2}}{p^{2}-\mu
^{2}+i\epsilon}.
\end{equation}

\section{Screening effects}
We show here the effects of massive fermion loop on the photon renormalization
constant and spectral function. The vacuum polarization of the four component
massive fermion is \cite{Bob}%
\begin{eqnarray}
\Pi(k) &  =&-\frac{e^{2}}{8\pi}[(\sqrt{-k^{2}}+\frac{4m^{2}}{\sqrt{-k^{2}}}%
)\ln(\frac{2m+\sqrt{-k^{2}}}{2m-\sqrt{-k^{2}}})-4m],\nonumber\\
&  =&-\frac{e^{2}}{8}i\sqrt{-k^{2}}(-k^{2}>0,m=0),\nonumber\\
&  =&\frac{e^{2}}{6\pi m}k^{2}+O(k^{4})(-k^{2}/m\ll1).
\end{eqnarray}
In the massless photon approximation, the effects of screening is the largest. We
get the photon spectral function by the imaginary part of the dressed photon
propagator%
\begin{equation}
\rho_{\gamma}(k^{2})=\frac{\mbox{Im}D_{F}(k)}{\pi}=\frac{1}{\pi
}\mbox{Im}\frac{1}{(-k^{2}+\Pi(k))}.
\end{equation}
The renormalization constant of the photon field is defined by%
\begin{equation}
Z_{3}^{-1}=\int_{0}^{\infty}\rho_{\gamma}(s)ds=1(m=0).
\end{equation}
There is a masless pole $\delta(s)$ for $m\neq0.$ But it does not contribute to
the fermion propagator. In that case, the fermion propagator vanishes in the limit
$\mu\rightarrow0.$ In numerical analysis, we set $m=c/N\pi,c=e^{2}N/8$ which are
expected in the  $1/N$ approximation.

\subsection{Quenched case}
The function $F$ is evaluated in the covariant gauge \cite{Hoshino}%
\begin{eqnarray}
F  &  =\frac{e^{2}}{8\pi}[\frac{\exp(-\mu|x|)-\mu|x|E_{1}(\mu|x|)}{\mu}%
-\frac{E_{1}(\mu|x|)}{m}\nonumber\\
&  +\frac{(d-1)\exp(-\mu\left\vert x\right\vert ))}{2\mu}],|x|=\sqrt{x^{2}}.
\end{eqnarray}
For short and long distance, we have the approximate form of the function $F$
\begin{equation}
F_{S}\sim A-\mu|x|+(D+C|x|)\ln(\mu\left\vert x\right\vert ))-\frac
{(d+1-2\gamma)e^{2}|x|}{16\pi},(\mu|x|\ll1).
\end{equation}%
\begin{equation}
F_{L}\sim0,(1\ll\mu|x|).
\end{equation}
From the above formulae, we have the approximate form of the spectral function
$\exp(F)$
\begin{equation}
\exp(F)=(%
\begin{array}
[c]{c}%
A(\mu|x|)^{D+C|x|}(\mu|x|\leq1)\\
1(1\leq\mu|x|)
\end{array}
),
\end{equation}
where $m$ is the physical mass and%
\begin{equation}
A=\exp(\frac{e^{2}(1+d)}{16\pi\mu}+\frac{e^{2}\gamma}{8\pi m}),C=\frac{e^{2}%
}{8\pi},D=\frac{e^{2}}{8\pi m},
\end{equation}
where $\gamma$ is the Euler's constant. From the above short distance behaviour
of $F$, we obtain the mass shift and the position dependent mass as%
\begin{equation}
\triangle m=\frac{e^{2}(1+d-2\gamma)}{16\pi}+\frac{e^{2}\mu}{8\pi
m},m(x)=m-C\ln(\mu\left\vert x\right\vert ),
\end{equation}
We see here that $D$ acts to change the power of $|x|$. For $D=1,S_{F}(0)$ is finite
and we have $\left\langle \overline{\psi}\psi\right\rangle \neq0.$ Thus if we
require $D=1,$ we obtain $m=e^{2}/8\pi.$

\subsection{Unquenched case}
In the unquenched case, we replace the quenched spectral function for the photon by the
dressed one
\begin{equation}
\exp(F(x))\rightarrow\int_{0}^{\infty}\rho_{\gamma}(s)\exp(F(x,s)ds.
\end{equation}
From now on, we fix the gauge $d=-1$ to avoid linear infrared
divergences. There exists a massless photon pole contribution to $\rho_{\gamma
}(s).$ However, if we take a limit $\mu\rightarrow0$, the fermion propagator
vanishes, which can be seen from eq(11). In Fig.1,2 we see the $N$ dependence of
screening effects on the renormalization constant $Z_{3}^{-1}$ and the
function $\exp(\widetilde{F}(x))$ respectively. From these figure we see that
the screening effects are large for small $N.$
\begin{figure}
\begin{center}
\includegraphics[width=13pc]{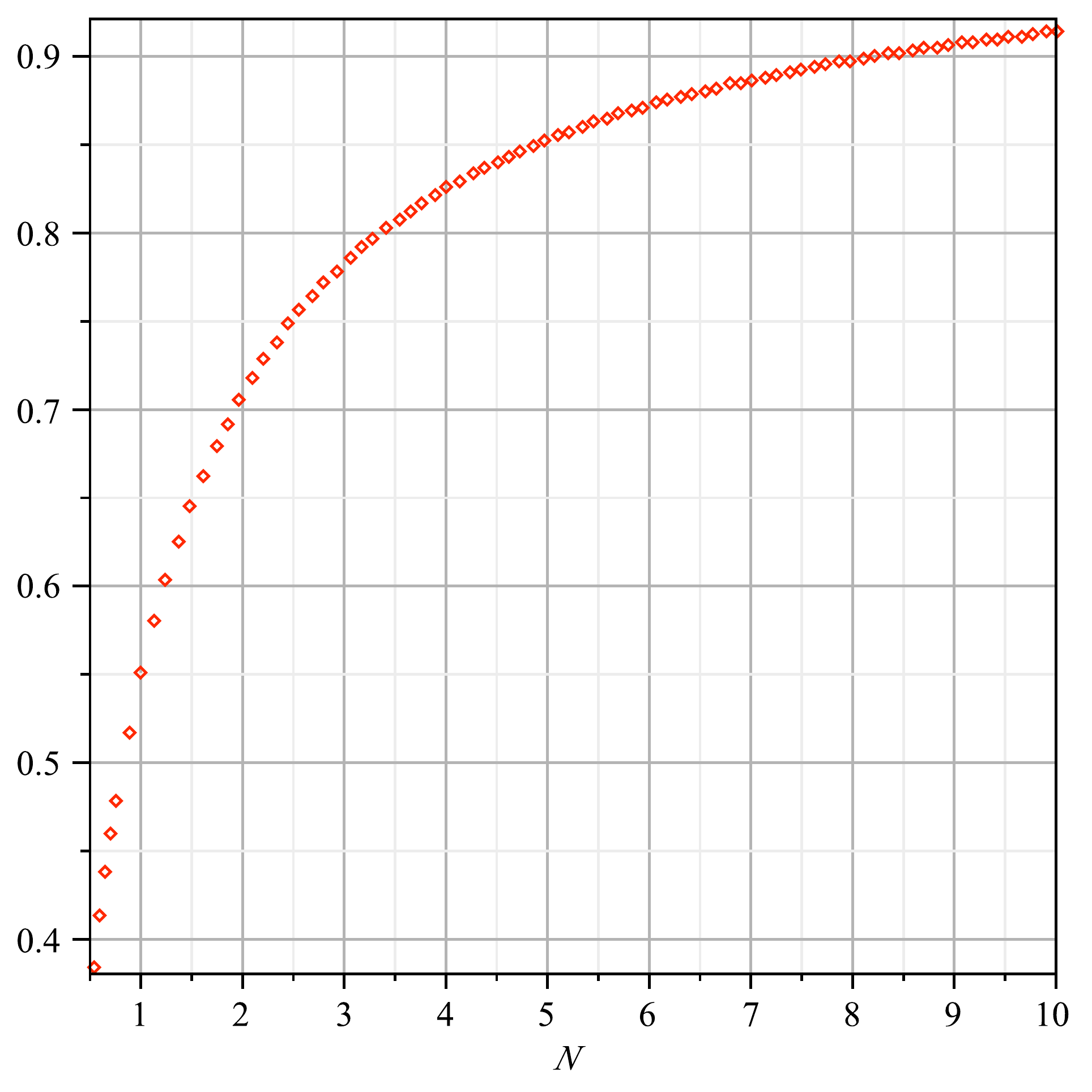}
\caption {$Z_{3}^{-1}-1$ for $m\neq0,N=1/2$ $..10.$}
\end{center}
\end{figure}
\begin{figure}
\begin{center}
\includegraphics[width=13pc]{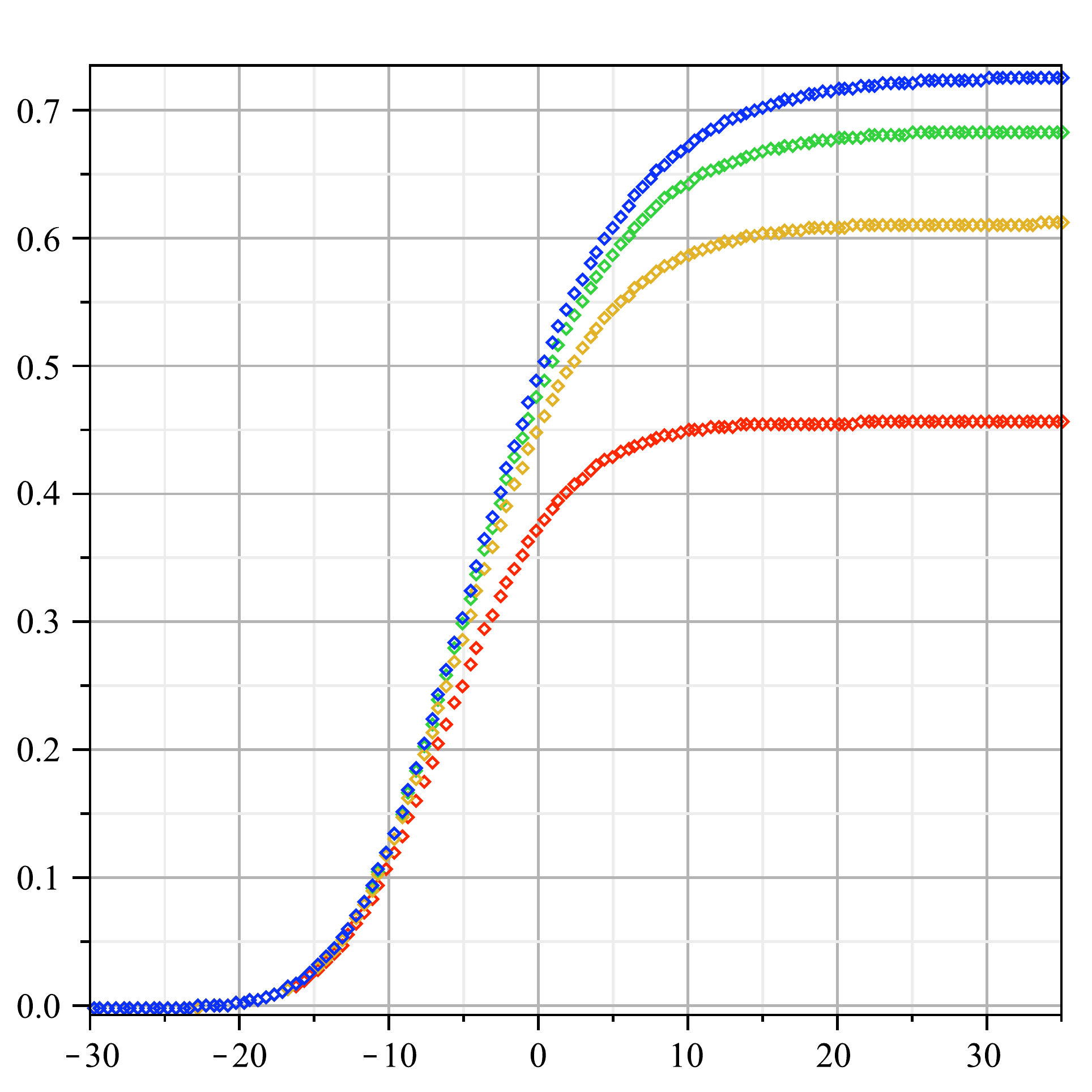}
\caption{$\exp(\widetilde{F}(x))$ for $N=1,2,3$ with $x=\exp(\pi/2\sinh(l/5)).$}
\end{center}
\end{figure}

\section{Minkowski space}
Here we change the variable $x^{2}\rightarrow iT,s=p^{2}/m^{2}$ \cite{Jackiw,Schwinger} and define
the spectral function of fermion $\rho(s)$ as
\begin{equation}
\exp(\widetilde{F}(iT))=\int_{0}^{\infty}d\sigma\rho^{F}(\sigma)\exp
(F(iT,\sqrt{\sigma})),
\end{equation}%
\begin{equation}
\rho(s)=\frac{1}{2\pi}\int_{-\infty}^{\infty}\exp(-i(s-1)T)\Im(\exp
(\widetilde{F}(iT))dT.
\end{equation}
In Fig.3, the spectral function $\rho(s)$, in unit of $e^{2}$, has sharp
peaks for both side of $s=1$ for $N=2,3$. For $N=1$ these peaks are wide. We see
that the real part of the propagator has no pole like singularity in
Fig.4, where $s=\ln(2/(1-\mbox{erf}(l/5))$. 
\begin{figure}
\begin{center}
\includegraphics[width=13pc]{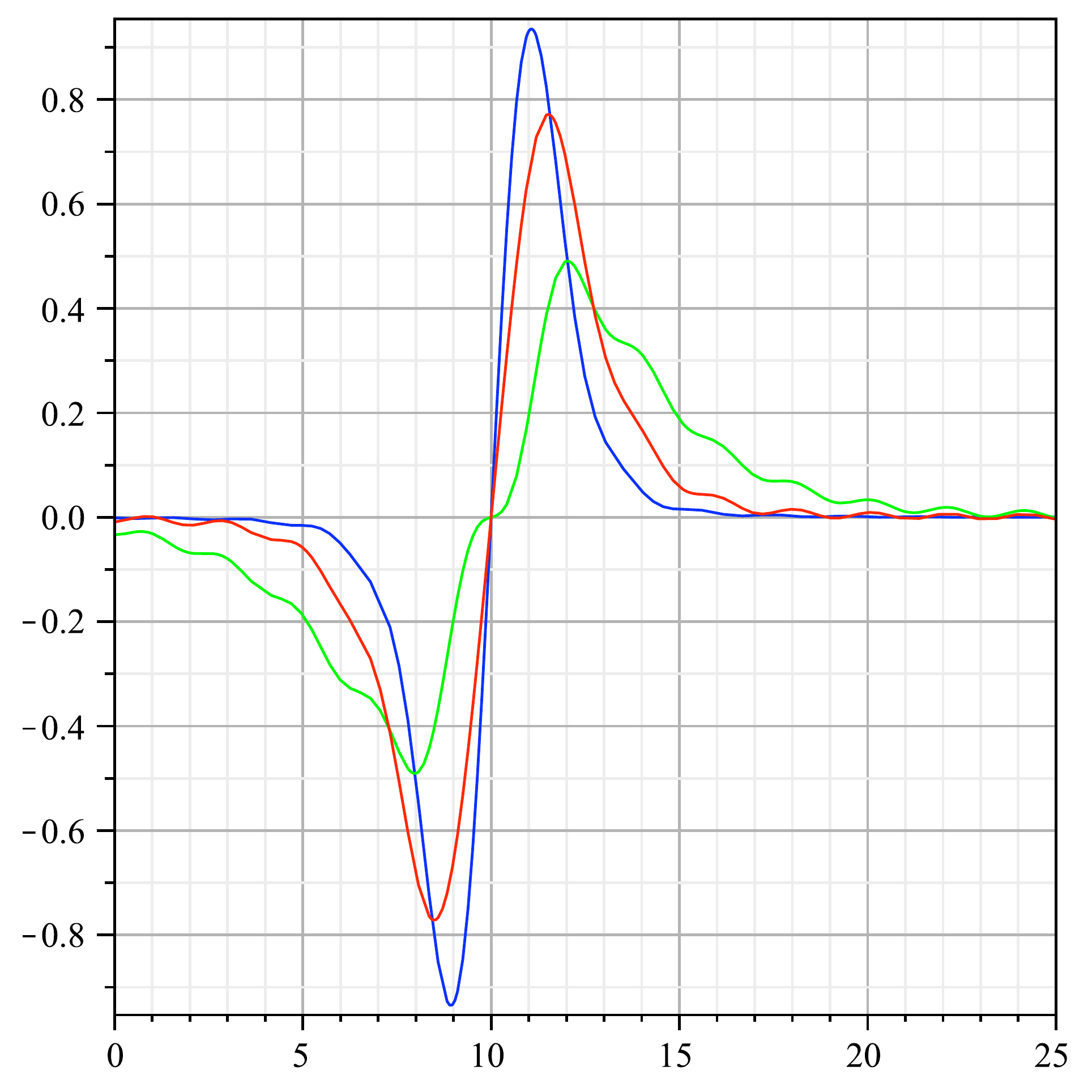}
\caption{$\rho(s)$ for $N=1(bottom),2,3(top)$ in unit of $e^2.$}
\end{center}
\end{figure}
\begin{figure}
\begin{center}
\includegraphics[width=13pc]{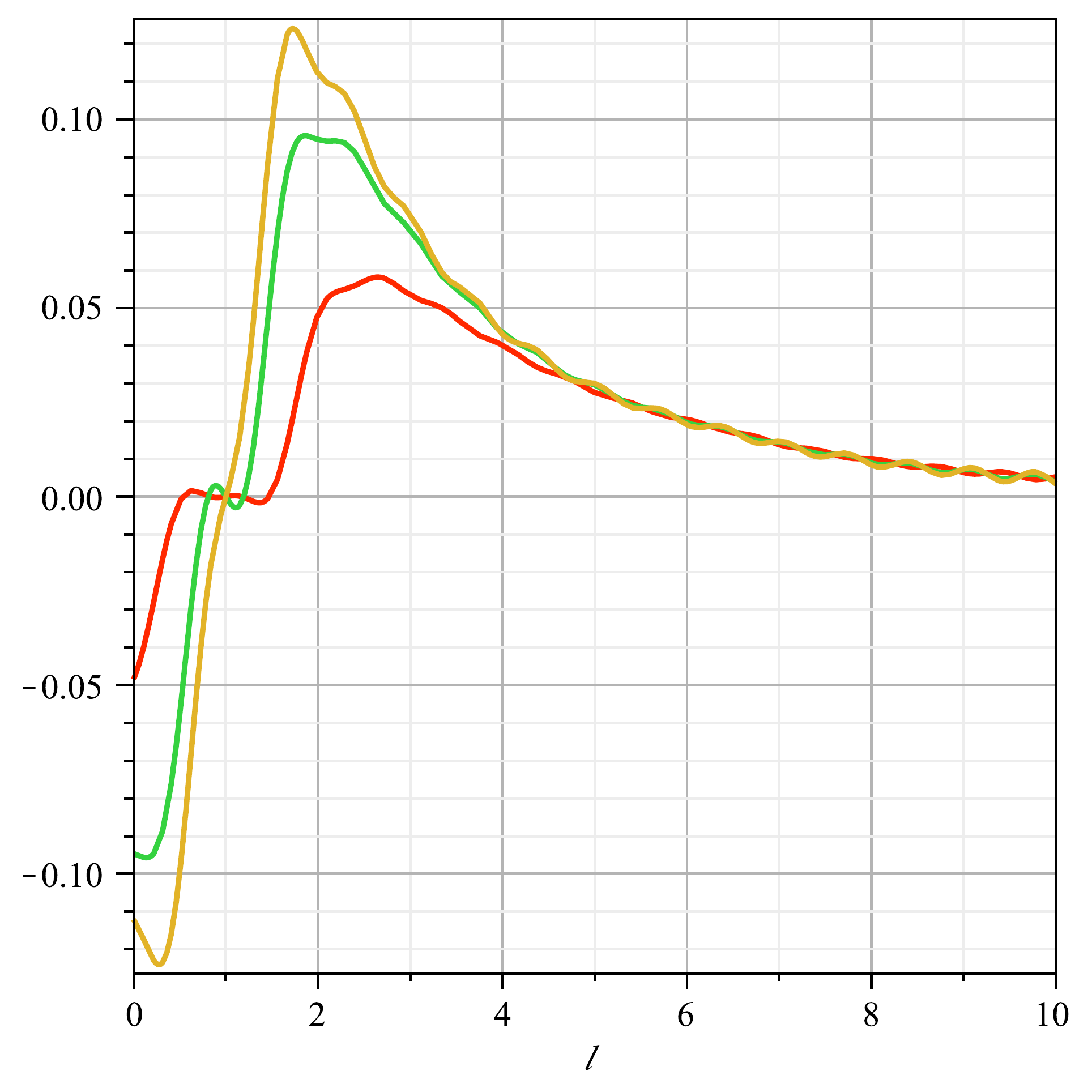}
\caption{$\Re(S_{S}(s)4\pi$ for $N=1(bottom),2,3(top)$ in unit of $e^{2}.$}
\end{center}
\end{figure}
For strong coupling, the spectral function has no peak which corresponds to the
double pole structure. This may be due to the large mass changing effects which
appeared as $c|x|\ln(\mu|x|)-c|x|.$

\section{Renormalization constant and order parameter}
In the beginning, we assume the asymptotic field :$\psi(x)_{t\rightarrow
+\infty,-\infty}\rightarrow\sqrt{Z_{2}}\psi(x)_{out,in}.$ The integral
representation of the propagator is written
\begin{equation}
S_{F}^{^{\prime}}(p)=\int_{m^{2}}^{\infty}\frac{(\gamma\cdot p+m)\rho
(s)ds}{p^{2}-s+i\epsilon}.
\end{equation}
However in our approximation we have%
\begin{equation}
Z_{2}^{-1}=\int_{0}^{\infty}\rho(s)ds=\int_{-\infty}^{\infty}dt\delta
_{+}(t)\mbox{Im}(\exp(F(it)))=0,
\end{equation}%
\begin{equation}
m_{0}Z_{2}^{-1}=0,
\end{equation}
for the case of positive anomalous dimention. Our approximation contains
the analytic solution for the linear D-S equation in the quenched Landau gauge \cite{Koopman}.%
\begin{figure}
\begin{center}
\includegraphics[width=13pc]{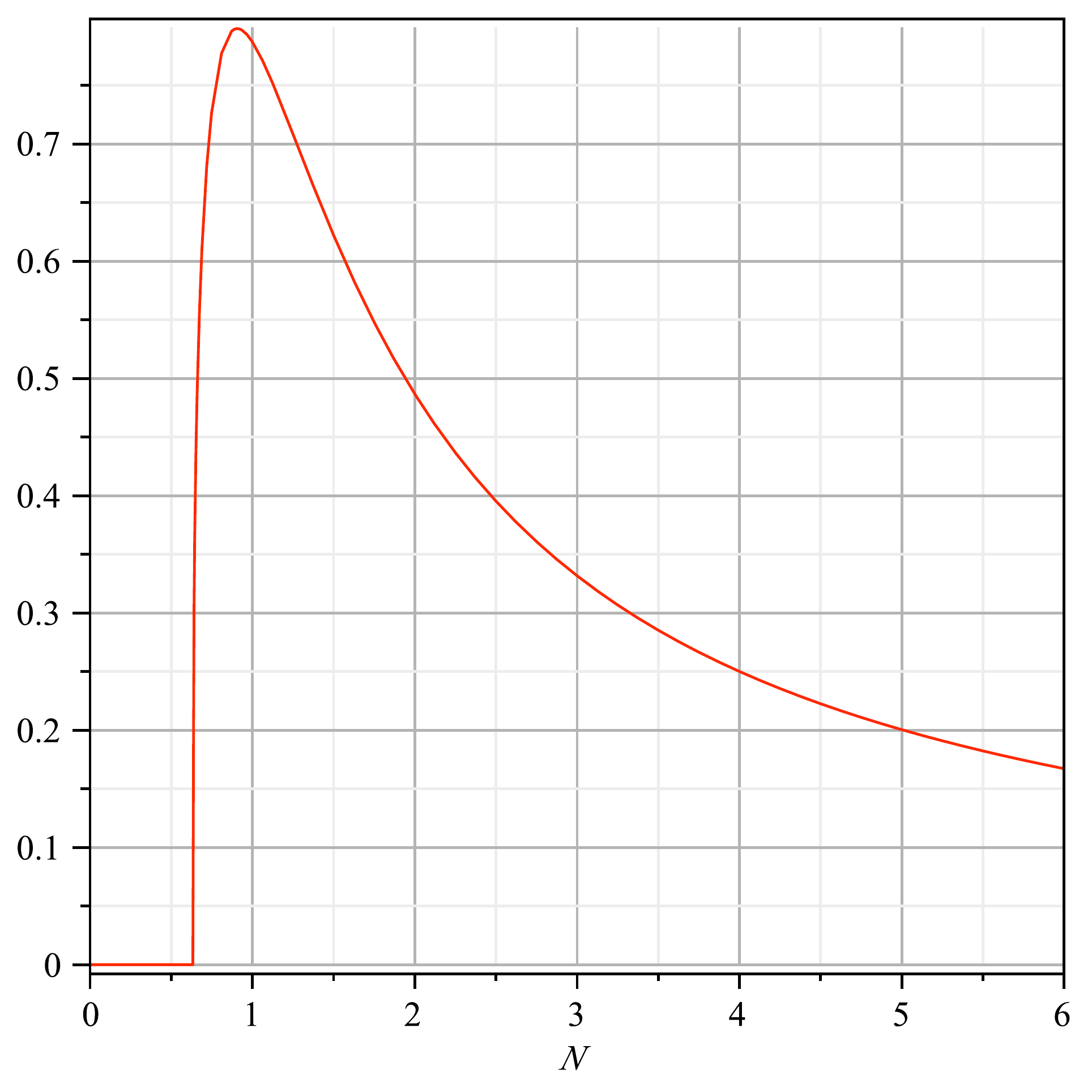}
\caption{$-\left\langle \overline{\psi}\psi\right\rangle $ in unit of $c=1.$}
\end{center}
\end{figure}
\begin{equation}
S_{S}(p)=\frac{m^{3}}{(p^{2}+m^{2})^{2}},\mbox{Im}S_{S}(p)=-m^{3}%
\delta^{\prime}(p^{2}-m^{2}).
\end{equation}
This behaviour is seen for weak coupling in our approximation. For
$D\mathbf{=}1,$ $m=c/N\pi$ we have $\left\langle \overline{\psi}
\psi\right\rangle =-4\lim_{x\rightarrow0}m\overline{\rho}(x),\left\langle
\overline{\psi}\psi\right\rangle \approx-(6.7-2.2)10^{-3}e^{2}$ for $N=1-3.$ In
the case of a massless fermion loop, the results are $(2.8-2.2)10^{-3}e^{2}$ for
$N=1-3$ in the same way. These values are consistent with those obtained
$(3.33,1.2)10^{-3}e^{2}$ for $N=0,1$ in D-S eq with massless fermion loop and
kinds of vertex ansatz \cite{Maris}. Finally we show the order
parameter $\left\langle \overline{\psi}\psi\right\rangle $ as a function of
$N$ and the coupling constant $c$. Here we notice that $\left\langle
\overline{\psi}\psi\right\rangle $ equals to zero for $N\leq2/\pi$ for
$c=e^{2}N/8=1$ in Fig.5. In this mass region, $m=c/N$ \ is large enough and
vacuum polarization is suppressed in such a way that $c/2m=N\pi/2\ll1$. In this
case, the photon propagator is the quenched one and $\left\langle \overline{\psi}
\psi\right\rangle $ vanishes. If we fix $N$ and vary the coupling constant
$c, $ large $c$ corresponds to the quenched case and there exists a critical
coupling $c_{r}=\pi N/2$ below which $\left\langle \overline{\psi}
\psi\right\rangle \neq0.$ This phenomenon is the same with finite temperature
phase transition in superconductivity. In the case of a massless fermion loop, there is only a broken phase from the above reasons.
\begin{figure}
\begin{center}
\includegraphics[width=13pc]{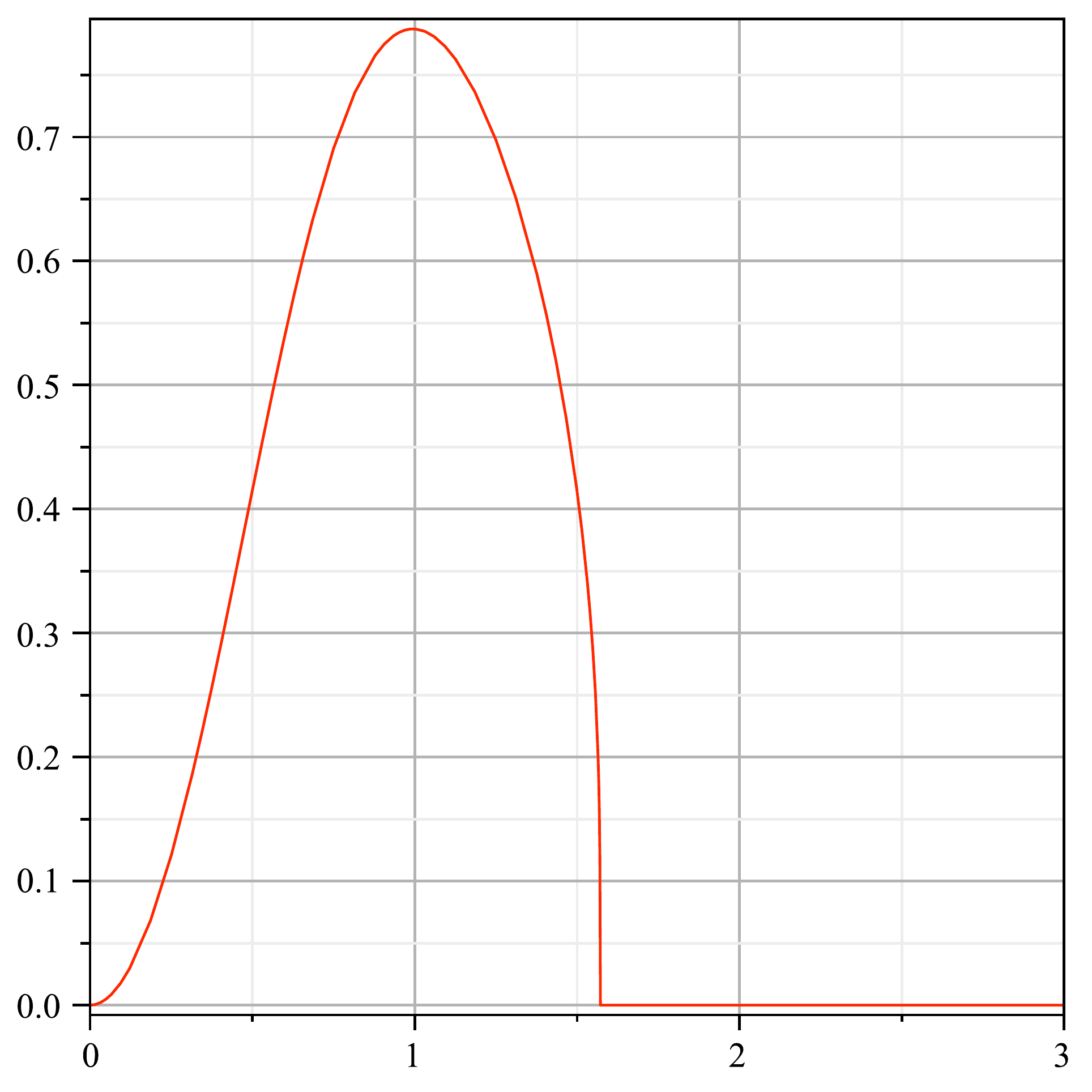}
\caption{$-\left\langle \overline{\psi}\psi\right\rangle (c)$ for $N=1.$}
\end{center}
\end{figure}
\section{Summary}
In this work, we examined whether the dispersion like method works or not to
determine the non perturbative fermion propagator in QED3. If we assume
physical mass, we have a mass shift, its log correction and anomalous dimension in
the lowest order spectral function. There is a gauge $d=-1$ in which linear
divergences vanish. Including vacuum polarization, the logarithmic infrared
divergence disappeared. As a results, there is no infrared divergences and no
poles. There exists a critical coupling constant such as critical temperature
in superconductivity. We have seen that the dispersion like method works
well in QED3. In the future, we will analyse finite temperature
case, QCD(2+1), and its application to high-T$_{c}$ superconductivity.

\end{document}